\documentstyle[prl,aps,12pt]{revtex}
\begin{document}
\draft
\renewcommand{\baselinestretch}{1.3}
\title{\Large\bf  Effective QED Actions: Representations,
 Gauge Invariance, Anomalies and Mass Expansions}
\author{S. Deser$^{(a)}$,   L. Griguolo$^{(b)}$  and 
        D. Seminara$^{(a)}$\footnote{e-mail:
{\tt Deser,Seminara@binah.cc.brandeis.edu; Griguolo@irene.mit.edu}}}
\address{\it $^{(a)}$ Department of  Physics, Brandeis University,
                        Waltham, MA 02254, USA\\
             $^{(b)}$ Center for Theoretical Physics, Laboratory
                         for  Nuclear Science and Department of Physics,\\
                         Massachusetts Institute of Technology, Cambridge,
                         Massachusetts 02139, U.S.A.}
\date{Received \today}
\maketitle
\medskip
\font\ninerm = cmr9
\pagestyle{empty}
\begin{abstract}
\ninerm\noindent
We analyze and give explicit representations for 
the effective abelian vector gauge field actions generated by
charged fermions with particular attention to the thermal regime in 
odd dimensions, where spectral asymmetry can be  present. We show, 
through $\zeta-$function regularization, that both small and  
large gauge invariances are preserved at any temperature
and for any number of fermions at the usual price of 
anomalies: helicity/parity invariance will be lost in even/odd 
dimensions, and in the latter even at zero mass. 
Gauge invariance dictates a very general 
``Fourier'' representation 
of the action in terms of the holonomies that carry the novel,
large gauge invariant, information. 
We show that large (unlike small) 
transformations and hence their Ward identities, are 
not perturbative order-preserving, and clarify the role of 
(properly redefined) Chern-Simons terms in this context.
From a powerful representation of the action in terms of massless heat 
kernels,  we are able to obtain rigorous gauge  invariant expansions,
for both small and large fermion masses,
of its separate parity even and odd parts in arbitrary dimension. 
The representation also displays  both the nonperturbative 
origin of a finite renormalization ambiguity, and its physical
resolution by requiring  decoupling at infinite mass.
Finally, we illustrate these general
results by explicit computation of the effective action for some
physical examples of field configurations in the three 
dimensional case, where our conclusions on finite temperature 
effects may have physical relevance.
Nonabelian results will be presented separately. 
\end{abstract}
\pacs{PACS numbers:\ \  11.10.Wx 11.15 11.30.Er 11.30RD\hfill BRX-TH-417}
\vfill
\newpage
\pagestyle{plain}
\renewcommand{\thesection}{\arabic{section}}
\setcounter{section}{0}
\renewcommand{\theequation}{\arabic{section}.\arabic{equation}}
\section{Introduction}
\setcounter{equation}{0}
Effective gauge field actions, induced by integrating out their sources,
play an essential role in physics.  Here we will study the result of
integrating out charged fermions minimally coupled to an abelian
vector potential, with emphasis on odd dimensions, especially D=3, 
and on the thermal regime in which topological considerations are both
essential and delicate.  The corresponding nonabelian analysis will
be presented subsequently \cite{00}. 
QED$_3$ models are interesting for a number of reasons:
From a theoretical point of view they provide 
fascinating examples of interrelations between quantum mechanics,
unusual gauge invariance, topology and discrete space-time symmetries
\cite{zero,3D}.
On more physical grounds, they are natural candidates for the  description
of planar 
phenomena in the condensed matter context \cite{CondMatt} or the high  
temperature regime  of four-dimensional models \cite{FiniteTemperature}.   

Many intriguing features  of  odd-dimensional  dynamics stem from the 
existence of the unconventional, parity-violating, but apparently  gauge
invariant and well-defined
Chern-Simons  (CS) term, which has its simplest --quadratic-- form in the
planar ($d=3$) case \cite{3D}, 
\begin{equation}
\label{ICS}
I_{CS}=
\int d^3 x \epsilon^{\lambda\mu\nu}
~ A_\lambda F_{\mu\nu}.  
\end{equation}
However, as we shall see,  $I_{CS}$ is neither gauge invariant
(which is in fact one essential reason for its interest!) nor is it
generically well-defined, but it can be ``improved"; it also does not
appear ``unaccompanied" in the effective actions.  Understanding
these points plays a pivotal role both in analyzing QED$_3$, as well
as incorporating correctly possible ``bare" CS terms that could be
present in a descent from the D=4 QED topological action
``theta" term $\int F \wedge F$.  In this connection, 
one must also come to terms with the proper
quantization requirements, stemming from their
gauge dependence, on the coefficients  of $I_{CS}$.
We will deal with $I_{CS}$ in sec. 2,
as part of a general analysis of the complete effective actions and
their gauge properties, extending work begun in \cite{us}.
We will then review why the perturbative
expansion of the effective action 
in the coupling constant is not invariant  under large (not contractable
to the identity) gauge transformations, thereby invalidating
the usual perturbative Ward
identity counting. Analyzing how gauge invariance
constrains the form of the full effective action in terms of its dependence
on the variables carrying the local and global degrees of freedom,
namely the field strength and the holonomy, 
will bring in the CS term (in its correct, ``improved'' guise) 
as the carrier of global  information.  

In sec. 3, we shall detail
the properties of the Dirac determinant in the rigorous framework of
 $\zeta-$function regularization with particular attention to the
delicate interplay between large gauge invariance and spectral
asymmetry. This analysis will make manifest the necessary clash
in odd/even dimensions between parity/helicity and gauge invariance.
In odd dimension an ``intrinsic'' parity anomaly, {\it i.e.} one 
nonvanishing even for massless (hence formally parity-conserving)
fermions, is generally present and it is identified with
the $\eta-$function, more precisely with $\eta(0)$ \cite{pop}. 
This quantity will be seen to be a discontinuous
gauge invariant functional of the fields,  containing--as a single 
unit--the CS action together with a non-local object given by the
index; while the former can be easily recovered in
a perturbative approach, the latter, being discontinuous, becomes manifest
only through a nonperturbative investigation of the Dirac determinant.
We shall also notice that the parity-violating part of the
effective action suffers from a sign ambiguity whose mathematical origin
stems from having to specify a choice of cut in the definition  
of the complex power of
the Dirac eigenvalues; physically, this is a finite regularization effect
which has its counterpart also in the perturbative regime.
The ambiguity can be fixed by requiring a vanishing effective
action, {\it i.e.}, ``decoupling", in the infinite fermion mass 
limit.  We then obtain an explicit ``spectral" representation of 
the action in terms of massless heat kernels.

In sec. 4, we use this representation to derive systematic 
(gauge-invariant) mass-expansions in both small and large mass regimes. 
These expansions, valid for any dimensions, may have  a wider 
applicability and so are given in some detail.

In sec. 5,  many of the general features encountered in the previous
sections are illustrated by explicit integration in presence
of some specific, physically non-trivial, gauge field configurations. 
This also provides a useful check of the more formal results
developed  in the earlier sections. 

Many results presented  here, such as large gauge invariance and mass 
expansions, can be shown to extend straightforwardly to the 
non-abelian context. These, as well
some  features intrinsic to the non-abelian case will be discussed
in \cite{00}  and also illustrated through explicit configuration
examples.

\section{d=3: Large gauge invariance, effective action, CS terms}

\setcounter{equation}{0}

In this section we shall focus for concreteness  on the important 
and illustrative
case of $d=3$, but much of the discussion is  general. 
Our 3-space has $S^1({\rm time})\times \Sigma^2$ topology,
$\Sigma^2$ being a compact Riemann 2-surface such as a
sphere $S^2$ or a torus $T^2$, depending on the desired
spatial boundary conditions. We work with a finite 2-volume
in order to avoid infrared divergences  associated with
the continuous spectrum in an open space. Most
considerations presented in this section apply naturally
to more familiar 3-spaces, such as 
the usual $S^1\times I\!\! R^2$ assumed in the perturbative approach.
However, compact $\Sigma^2$ allowing for magnetic flux are more
physical and will become essential in our full nonperturbative 
construction below.

The $S^1$ circle is identified with  euclidean time and
its length $\beta=\displaystyle{{1}/{\kappa T}}$ is the
inverse of the temperature. Spinors are required to
satisfy antiperiodic boundary conditions, 
\begin{equation}
\label{abebound1}
\psi(t+\beta,{\bf x})=-\psi(t,{\bf x}),\ \ \
\ \ \ \ \bar\psi(t+\beta,{\bf x})=-\bar\psi(t,{\bf x}),
\end{equation}
while the $U(1)$ gauge field is chosen to be periodic,
\begin{equation}
\label{abebound2}
A_\mu(t+\beta,{\bf x})=A_\mu(t,{\bf x}).
\end{equation}
[In the presence of other non-trivial cycles,
such as $T^3$, one must specify the  periodicity
conditions also in their characteristic directions.] The fermion action
is taken to be 
\begin{equation}
\label{U1action}
S_f=i\int d^3 x   \bar\psi  \left (D\!\!\!\!/+m\right)\psi,
\end{equation}
where $D_\mu\equiv \partial_\mu + i A_\mu$ is the usual $U(1)$
covariant derivative, and the $\gamma^\mu$ are hermitian. 
Requiring  the gauge transformations $U$
\begin{equation}
\label{travo}
A_\mu\to A_\mu-i U^{-1}\partial_\mu U, \,\,\,\,\,\,
U\equiv\exp\left(i\Omega(t,{\bf x})\right), 
\end{equation}
to respect these periodicities  forces them to be periodic as well,
but allowing the phase $\Omega$ to obey 
\begin{equation}
\label{cippi}
\Omega(t+\beta,{\bf x})-\Omega(t,{\bf x})=2\pi n,\
\ \ n\in Z\!\!\! Z.
\end{equation}
Different $n$ in  (\ref{cippi}) specify gauge transformations
belonging to different homotopy classes; only transformations with the 
same $n$ can be continuously
deformed into each other. Those  $\Omega(t,{\bf x})$ with $n\ne 0$
generate ``large" gauge transformations. A representative
for each such class can easily be constructed,
\begin{equation}
\label{Un}
U_n(t,{\bf x})=\exp\left (i\frac{2\pi}{\beta} n t\right).
\end{equation}
[The composition law $U_n\times U_m=U_{n+m}$ expresses the mathematical
statement that $\Pi_1(U(1))=Z\!\!\!Z$.] Understanding how the invariance under
the transformations (\ref{Un}) constrains the form of the effective
action is a central issue. We  begin
by showing that the existence of a nontrivial $U_n$  invalidates 
the usual perturbative Ward identity  counting.  Restoring  (for the
moment) explicit  dependence on the coupling constant $e$, a gauge
transformation has the form
\begin{equation}
\label{transf}
A_\mu\to A_\mu-\frac{i}{e} U(t,{\bf x})^{-1}\partial_\mu U(t,{\bf x})=
A_\mu\to A_\mu+\frac{1}{e}\partial_\mu\tilde \Omega(t,{\bf x}),
\end{equation}
with $U(t,{\bf x})=\exp( i\tilde \Omega(t,{\bf x}))$. If $n=0$, 
$\tilde\Omega(t,{\bf x})$ is strictly periodic (``small" transformation),
and the apparent non-analytic $1/e$ behavior   in  (\ref{transf}) can be
made to disappear by redefining $\tilde \Omega(t,{\bf x})=e\Omega(t,{\bf
x})$. Thus a perturbative expansion will be small-invariant order by
order because, after the rescaling,  (\ref{transf}) cannot mix
different orders of perturbation theory. Instead under the { large}
transformations (\ref{Un}),
the gauge connection changes as follows
\begin{equation}
\label{shift}
A_0\to A_0+\frac{2\pi}{e} n \ \ \ \ \  A_i\to A_i.
\end{equation}
A rescaling will merely hide the $1/e$ factor in the boundary conditions,
leaving (\ref{shift}) unaffected. This intrinsic $1/e$ dependence means
that only the {\it full} effective action (as we shall show), but not its
individual expansion terms (including CS parts !) will remain 
large gauge invariant.
In fact the shift in  (\ref{shift}) can mix all orders of perturbation
theory. [Perturbative non-invariance  will also characterize 
any  other expansion that fails to commute  with the above boundary
condition.]

Let us now see precisely how {\it large} gauge invariance 
restricts  the possible structure of the determinant,
or indeed of any well-defined gauge field functional.  To simplify 
our argument and avoid irrelevant spatial homotopies, we shall take 
$\Sigma^2$ to be the sphere.  Because of the existence of the non-trivial 
$S^1$ cycle, we can construct, 
besides $F_{\mu\nu}$, a second and independent gauge invariant  object,
the holonomy: 
\begin{equation}
\label{bulla}
W(\beta, {\bf x})
\equiv\exp\left( i\int^\beta_0 A_0(t^{'},{\bf x})~dt^{'}\right )
\equiv\exp(i\beta {\cal A}_0({\bf x})) . 
\end{equation} 
We will show 
that $F_{\mu\nu}$ and $W$ together completely specify 
$A_\mu$ up to a gauge.
What information carried by $W(\beta,{\bf x})$, or equivalently
by ${\cal A}_0$,
is not already contained in the field strength? The gradient of $W$ obeys
\begin{equation}
\label{coito}
{\bf \nabla} W = 
iW \int^\beta_0 dt^{\prime} 
[\nabla A_0 (t^{\prime}, {\bf x}) -
\partial_{t^{\prime}} {\bf A} (t^{\prime}, {\bf x})] =
-iW\int^\beta_0~{\bf E}(t^{\prime},{\bf x})~dt^{\prime}, 
\;\; E_i \equiv F_{0i}
\end{equation}
because $\displaystyle\int^\beta_0 dt^\prime \partial_{t^\prime} {\bf A}
(t^\prime,{\bf x})$ vanishes by  periodicity (\ref{abebound2}); equivalently
\begin{equation}
\label{coito1}
-{\bf \nabla} {\cal A}_0({\bf x})=\frac{1}{\beta}\int^\beta_0 dt^{\prime}~
{\bf E}(t^{\prime},{\bf x}).
\end{equation}
Since $W$ is unimodular, the linearity of (\ref{coito})  
implies that $W$ is the  product of a  (generically nonlocal) functional of 
${\bf E}$  and of the 2-geometry times a constant  phase factor. 
The integrability
(vanishing curl) of (\ref{coito}) is insured by the Bianchi identity; its
general solution is obtained from the divergence of (\ref{coito1}) to
yield
\begin{equation}
\label{coito2}
{\cal A}_0({\bf x})=\frac{2\pi}{\beta} a -
\int^\beta_0dt^{\prime}~ \int d^2y G({\bf x,y}){\bf \nabla}
\cdot {\bf E}(t^{\prime},{\bf y}).
\end{equation}
Here  the Green's function $G({\bf x,y})$ on the two-sphere obeys
$\triangle G({\bf x,y})=1\!\!1- I\!\!P$, where $I\!\!P$ is the projector on
the zero-modes. The constant part of
${\cal A}_0({\bf x})$ corresponds to
the constant phase part $\exp(2\pi i a)$ of $W$,
while the rest of ${\cal A}_0({\bf x})$ exhibits the ${\bf E}$ and
$2-$geometry dependence. 
Thus the  new information carried by $W$ is encoded entirely in 
the topological degree of freedom  $a$, the flat 
connection;\footnote{The appearance of topological degrees of 
freedom governing  behavior
under homotopically non-trivial transformations is not unusual and occurs
in other contexts and dimensions. In two dimensions, for example,
all the dynamics of Yang-Mills theories is described by such variables.}
it transforms  according to $a\to a+1$ under { large} transformations. 
The fermion 
determinant can now be viewed as a functional of both $F_{\mu\nu}$
and $a$. Its invariance is assured if the effective action $\Gamma$  obeys the 
additional {\it finite} Ward identity 
\begin{equation}
\label{ZN}
\Gamma(a+1, F_{\mu\nu})=\Gamma(a,F_{\mu\nu}),
\end{equation}
namely if $\Gamma$ is periodic in $a$:  equivalently,
(\ref{ZN})  expresses the invariance of $\Gamma$ under the abelian
large  transformation group 
$Z\!\!\! Z$.  The periodicity in $a$ permits us to 
Fourier-expand $\Gamma$:
\begin{equation}
\label{coppola0}
\exp\left( -\Gamma(F_{\mu\nu},a)\right)=\sum_{k=-\infty}^{\infty}
\left(\hat\Gamma^{(1)}_k (F_{\mu\nu})\cos2\pi k a+\hat\Gamma^{(2)}_k
(F_{\mu\nu})\sin 2\pi k a \right)
\end{equation}

Before going further, however, we want to reexpress $a$ in terms of an
appropriate functional of the gauge field (but of course not of
$F_{\mu\nu}$  alone, as it is insensitive to $a$).  This is precisely
the role of $I_{CS}$, as defined by (1), 
(or rather of its corrected version, as we shall see) and we must
therefore consider its properties in detail.  Under the gauge transformation
(9), we have
\begin{equation}
\label{abelvar}
I_{CS}(A^{U})\longrightarrow I_{CS}(A)+
\int d^3 x 
\epsilon^{\lambda\mu\nu}~\partial_\lambda\left(\Omega F_{\mu\nu}\right)
= I_{CS} (A) + \Delta I_{CS}. 
\end{equation}
Although the gauge  term in (\ref{abelvar}) is a total divergence, dropping
it  is not generally permitted, as we now see: 
\begin{eqnarray}
\label{abelvar1}
\Delta I_{CS}&=&\int d^3 x\epsilon^{\lambda\mu\nu}~\partial_\lambda\left
[\Omega(t,{\bf x})F_{\mu\nu}\right ]=2\int^\beta_0 dt\partial_t \int_{\Sigma} 
d^2x\left[\Omega(t,{\bf x}) B(t,{\bf x})\right ]\nonumber\\
&+&2\int^\beta_0 dt \int_{\Sigma}
d^2x\epsilon^{ij}
\partial_i \left [\Omega(t,{\bf x}) E_j\right]
=2\int_{\Sigma} d^2 x\left[\Omega(\beta,{\bf x})-
\Omega(0,{\bf x})\right] B(0,{\bf x}).
\end{eqnarray} 
The magnetic ($B\equiv F_{12}$)  and electric fields, 
being physical, must be periodic in  $t$. 
The  electric contribution 
in (\ref{abelvar1}) vanishes if  $\Sigma^2$ does not allow non-trivial boundary 
conditions (the gauge invariant $E_i$ cannot have jumps, while $\Omega$ 
must also be a well-defined 2-scalar on $\Sigma^2$). 
On the other hand, the magnetic term does not vanish for large
transformations, where (\ref{abelvar1})  becomes
\begin{equation}
\label{Fluxo}
\Delta I_{CS}=
4\pi n\int_{\Sigma^2} d^2 x B(0,{\bf x})=4\pi n\Phi_{\Sigma^2}(B).
\end{equation}
The magnetic flux $\Phi$ is in general non-vanishing, time-independent
(by the Bianchi identity) and, as a topological necessity \cite{Naka}, 
$\Phi /2\pi$ is integer quantized.
It would thus seem that  any bare CS action, 
conveniently defined as $\mu/16\pi^2 I_{CS}$, shifts by
${\mu k n}/2$ under large gauge changes.
Consequently, the requirement that the phase exponential of any action
(the relevant object at the quantum level) be gauge invariant 
would seem to enforce
the quantization condition\footnote{Mathematically
this quantization relies on the fact that $\Pi_1 (U(1))= Z\!\!\!
Z$.  In the non-abelian regime, quantization of $\mu/2\pi$ is  of course
always required \cite{3D}.} on the CS parameter $\mu/2\pi$
that it be an even integer. 
Unfortunately, while this quantization argument 
is attractive, $I_{CS}$ 
is not even well-defined, precisely due to the very reason, 
$\Phi \ne 0$ for quantization!  Briefly stated,
$\Phi \ne 0$ requires non-trivial connections ${\bf A}$
on $\Sigma^2$, thereby making $I_{CS}$ manifestly patch-dependent 
(a well-defined action
is not patch-dependent!). This major deficiency in $I_{CS}$ should make
one suspicious of the validity of the above quantization requirement.
Fortunately, $I_{CS}$ can be ``improved",
but we will see that the quantization of the (bare) coefficient
of the ``improved''  $I_{CS}$ becomes $\mu/2\pi= n$ rather than $2n$.

We now sketch a heuristic ``derivation" of $I_{CS}$, since precise ones
were given long ago \cite{6b,6c,WD}; a new derivation \cite{Semi} will
also justify it.  Consider the particular gauge transformation
\begin{equation}
\Omega (t,{\bf x})=-\left [\int_0^t  dt^\prime-\frac{t}{\beta}
\int_0^\beta dt^\prime \right ]
A_0 (t^\prime,{\bf x})
\equiv {\cal O} A_0 \equiv \tilde{A}_0 (t,{\bf x}),
\end{equation}
in its effect on $I_{CS}$.  
Since $\Omega (t,{\bf x})$ is manifestly periodic in $t$
($\Omega (\beta ) = \Omega (0) = 0$), we are allowed to 
neglect the  divergence in ($\Delta I_{CS}$) 
and thus $I_{CS} (A)=I_{CS} (A^U)$,
where it is easy to check that the transformed fields are\footnote{Note
that the ${\cal O}$ operation projects out any time-independent factors.}
\begin{equation}
A^U_0(t,{\bf x})=\frac{1}{\beta}
\int^\beta_0 dt^\prime A_0(t^\prime,{\bf x})={\cal A}_0 ({\bf x}),\
\ \ A^U_i(t,{\bf x})=A_i(0,{\bf x})+{\tilde E}_i (t,{\bf x}).
\end{equation}
In terms of these variables, $I_{CS}$ has the form
\begin{eqnarray}
\label{cip44}
I_{CS} (A)
&=&2\int^\beta_0 d t \int d^2 x \left[ {\cal A}_0({\bf x}) B(t,{\bf x})+ 
\epsilon^{ij} ({\tilde E}_i(t,{\bf x}) 
+ A_i(0,{\bf x})) E_j (t,{\bf x})\right ]=\nonumber\\
&=&2\int^\beta_0 d t \int d^2 x \left( {\cal A}_0({\bf x}) B(t,{\bf x})+ 
\epsilon^{ij} {\tilde E}_i(t,{\bf x}) E_j(t,{\bf x})
+ \epsilon^{ij}A_i(0,{\bf x}) \partial_j {\cal A}_0({\bf x}) \right )=\nonumber\\
&=&2\int^\beta_0 d t \int d^2x \left [ {\cal A}_0({\bf x}) (B(t,{\bf x})
+B(0,{\bf x}))+ \epsilon^{ij} {\tilde E}_i(t,{\bf x}) E_j(t,{\bf x}) \right ]-
K,\\
K&\equiv& -2\int d^3x\epsilon^{ij} \partial_j [A_i(0,{\bf x}) {\cal A}_0
({\bf x})]
\end{eqnarray}
where, to reach the last term of the second equality, we have used $E_j(t,{\bf x})=
-\partial_j
{\cal A}_0({\bf x})+\partial_0 A^U_j(t,{\bf x})$ and then dropped  
$\partial_0 {A}^U_j(t,{\bf x})$ by periodicity. The sum $\bar I_{CS}\equiv
I_{CS}+K$ is perfectly well-defined (and small gauge invariant) 
since it contains no explicit ${\bf A}$ dependence
and represents the
advertised ``improved" CS term. The boundary term $K$ fails to vanish
for interesting configurations, involving non-trivial flux $\Phi$, because
there the connection ${\bf A}$ is different on the two patches that cover
the sphere.\footnote{The patch dependence of $K$ is easily
described schematically: Consider two patches defined by 
(for simplicity) some
arbitrary latitude cut $\theta = \theta_0$.  Then if $A^\pm$ denote
the respective values of the $A_\phi (\theta_0)$ on the upper/lower
caps, it is manifest that $K \sim \int d\phi (A^+ - A^-){\cal A}_0
(\theta_0)$.  Clearly, $K$  depends on the patch choice $\theta_0$ and
does not vanish if $\Phi \neq 0$, due to the usual non-trivial gauge
gauge difference on the patches familiar from magnetic monopole
constructions.} 
[Note that  perturbative calculations in the usual expansion 
about a trivial ($A_\mu=0$) connection
will never see the $K$ term; to include a reference background
would complicate even the one-loop computation considerably.]  We may now
rewrite $I_{CS}$ as the sum of terms depending only on the 
$F_{\mu\nu}$ together with those depending on the flat connection,
from the constant part\footnote{Here another arbitrary choice was made in
keeping the constant part of ${\cal A}$ even though it appeared in differentiated
form in the second equality's last term.  This choice of what
{\it physical} term to divide between $K$ and $\bar{I}_{CS}$ led to the
coefficient shown in (24).  Fortunately, this is also the correct answer from
\cite{6b,6c,WD,Semi}.}  of  ${\cal A}_0$ in (17),
\begin{equation}
\label{coop}
\bar I_{CS}= 8\pi a\Phi +Q[F]
= 16\pi^2 na + Q(F), \;\;\;\;
n \epsilon Z\!\!\!Z \; .
\end{equation}
This representation thus demonstrates that $\bar I_{CS}$ is not 
independent, but is
determined by $a$ and $F$,  its behavior under large transformations being
completely governed by $a$. It also enables us to compute the correct
quantization of the coefficient in a {\it bare} $\bar I_{CS}$
action (which must of course depend on $\bar I_{CS}$, not $I_{CS}$!): Under
a large gauge transformation, $a\to a+1$, $\bar I_{CS}\to\bar I_{CS}+
8\pi\Phi$, that is $\bar I_{CS}$ changes by $16\pi^2 k$, so that the
bare $\mu/2\pi$ must be an integer, not just an even one.

Having established the role of $\bar{I}_{CS}$ as the carrier of the 
holonomy information, we return to the Fourier expansion of the
action (\ref{coppola0}) and reexpress the $a$-dependence there in terms of
$\bar{I}_{CS}$
\begin{equation}
e^{-\Gamma (F,\bar{I}_{CS})} =
\sum^\infty_{k=-\infty} 
\left[ \Gamma^{(1)}_k (F) \cos k(\bar{I}_{CS}/8\pi n)
+\Gamma^{(2)}_k (F) \sin k(\bar{I}_{CS}/8\pi n) \right] .
\end{equation}
This form will be concretely realized by explicit field configurations in
sec. 5.  For our purposes, it shows how explicit CS terms can be
present, when ``protected" by sines and cosines, without loss of
large invariance, but this invariance is lost in a power series expansion.
As is necessary, we will confirm
this formal analysis in sec. 3, when we obtain
the properly regularized determinant.

We are now in a position to settle and old paradox arising in naive
perturbative calculations of $\Gamma$:  At one-loop  
(which is everything if the photon is not dynamic) level, the fermions give 
rise to an effective CS contribution, irrespective of whether there is an
initial bare one. The calculation of the coefficient is straightforward
\cite{loopT},   
\begin{equation}
\label{DeltamT}
\frac{\Delta\mu}{2\pi}=\frac{e^2}{2} {\rm tanh}\left[\frac{\beta m}{2}\right],
\end{equation}
which is noninteger for generic $\beta=(\kappa T)^{-1}$. However, since at the same time 
it was (correctly) thought both that (6) seems to
signal an irremediable large gauge anomaly and that the matter action
$\displaystyle{I_f=i\int (dx)\bar\psi({D\!\!\!\!/}+m)\psi}$
and the process of integrating out
its excitations to obtain the effective action 
$\exp(-\Gamma[A])=\det(i{D\!\!\!\!/}+i m)$
should be intrinsically gauge invariant, this  paradox 
has generated a considerable literature.
Opinions  have differed widely: one claim is
that there is no anomaly, due to some "obscure" non-perturbative mechanism 
that will restore the integer nature of $\mu$ \cite{Pisarski,Rossini(a)}.
Specifically \cite{Pisarski}  conjectured that the usual
perturbative definition of the CS coefficient through the two-point
function is not physically relevant and a possible nonperturbative one
in terms of the complete effective action was proposed. Given a large gauge
transformation $U_L$ of winding number $n$, a new renormalized $\mu_R$ is
to be defined according to 
$2\pi n \mu_R =\Gamma[U_L^{-1}\partial_\mu U_L]-\Gamma[0],$
and its integer nature 
is supposed to be protected by some topological Ward identity. In
\cite{Rossini(a)}, under the (incorrect) assumption that the only
parity violating contribution in the effective action is the CS term,
it is shown that the path-integral formulation of the theory is consistent
only if $\mu$ remains a temperature-independent integer. Another point
of view accepts the temperature dependence in (\ref{DeltamT}) as a correct
prediction of the theory entailing, for example, the breakdown of the
anyonic description of superfluidity \cite{CondMatt}. 

We have already seen how to dispel the paradox formally.  A
first step in understanding the real nature of this puzzle was recently
taken in \cite{Dunne}; a solvable one-dimensional abelian analog
of the problem was carefully analyzed and in particular its
effective action was computed in closed form: While gauge invariant 
(at least for an even number of fermions), its
perturbative expansion indeed contained a (one-dimensional) CS term with the
temperature-dependent coefficient  (\ref{DeltamT}). This result
thus allowed the coexistence of large gauge invariance of the full action
and non-quantization of the perturbative CS coefficient.
It was then established in  \cite{us}, that the effective action, 
independent of the
number of fermions, is indeed invariant under both small and large
transformations using the classic results of \cite{Gilkey,Gilkey2} 
that permit a
clear definition of  the Dirac operator's functional determinant 
by means of $\zeta-$function regularization, as we
shall show in detail in sec. 3.  We shall also see how Chern-Simons term's
noninvariance is precisely  compensated by accompanying
non-local contributions
in the effective action that are not perturbatively visible.

Finally we mention another historical misunderstanding
which goes back  to the original  papers, \cite{Red} and \cite{witten}:
the relation between the number of fermions and gauge invariance
in three dimensions. 
It is often stated that, in complete analogy
with the $SU(2)$ anomaly in four dimensions \cite{SU(2)},
large gauge invariance in three dimensions is
maintained only for an even number of fermions or more precisely for a
certain choice of matter multiplets \cite{Dunne,SUSY}. 
What is true here is that
in the even $N_f$ case one can define  somewhat different regularizations
that preserve both gauge and parity, something that  is indeed not achievable
for odd $N_f$. With our regularization prescription, however, large gauge 
invariance is always preserved, while parity is always anomalous
 for both  even and odd number of charges.

\section{The~Action:~Regularization,~Representation~and~Anomalies}

\setcounter{equation}{0}

We now turn to the implementation of the formal framework of sec. 2,
by regularizing the fermion determinant and then exhibiting its properties.
We shall review the definition of the
Dirac operator's determinant in the rigorous framework
of the $\zeta-$function approach \cite{Gilkey,Gilkey2,Hawking} for arbitrary
dimension. Although this has become a very popular technique and a
well-established mathematical subject, we believe  it 
is worth reexamining in order to  point out some subtleties   
peculiar to odd-dimensional manifolds. 
Specifically,  we will  stress the delicate interplay between 
spectral asymmetry, large gauge invariance, parity anomalies
and perturbative expansions.
In the process a compact integral representation of
the $\zeta-$function for massive electrons 
in terms of the massless gauge invariant heat-kernels
will be derived for all dimensions. It will enables us to provide,
in sec. 4,  detailed  expansion of both the parity odd/even parts
for small and large fermion masses.

The mathematical tool that allows us to make sense of the formal 
product of the eigenvalues, $\displaystyle{\prod_{\lambda_n} \lambda_n}$,
defining the determinant is  $\zeta-$function regularization, which,
for  normal  operators such as $i({D\!\!\!\!/}+m)$ on a compact
manifold, reduces 
to
\begin{equation}
\label{zeta}
\zeta(s)\equiv\sum_{\lambda_n\in\ {\rm Spectrum}} (\lambda_n)^{-s};
\end{equation}
in the sum each eigenvalue $\lambda_n$ in the spectrum is  repeated according 
to its multiplicity.\footnote{There is an intrinsic ambiguity, the scale 
dependence of the  dimensionful $\lambda_n$, hidden in (\ref{zeta}). 
Strictly speaking, to construct
the $\zeta-$function one should use the dimensionless ratio $\lambda_n/
\mu$, with an arbitrary scale $\mu$.  The determinant is therefore
actually undetermined up to terms proportional to
$\zeta(0)\log\mu$ \cite{Hawking},
namely to the well-known trace anomaly. In odd dimensions
this  contribution of course vanishes.  Note also that the extension to $N$
fermions simply involves the product of the individual determinants.}
The convergence of the series (\ref{zeta}) for 
${\rm Re}~s>d$ in $d$ dimensions is assured by a classical
result on the asymptotic growth of eigenvalues \cite{Gilkey2}, which
for the massive Dirac operator reads
\begin{equation}
\label{stima} 
\lim_{n\to\infty}~ n |\lambda_n|^{-d}\simeq \rm const.
\end{equation}
Here  the   eigenvalue sequence  is ordered
so that $|\lambda_0|\le|\lambda_1|\le\cdots$. 
Actually one can go further and show that $\zeta(s)$ for $s>d$ defines
an analytic function that  can be extended to a meromorphic function with
only simple poles. In particular its analytic extension is regular at
$s=0$ and its derivative
there defines the determinant according to  Hawking's relation\footnote{
Although our discussion is focused on the Dirac operator, all the results
extend, with slight modifications, to the larger class of the elliptic
pseudo-differential operators \cite{Gilkey}  with a ray of minimal growth
(Agmon ray).} \cite{Hawking}
\begin{equation}
\label{determinante}
\det i(D\!\!\!\!/+m)=\exp(-\zeta^\prime(0)); \ \ \
\Gamma[A]=\zeta^\prime (0).
\end{equation}
Since the complex power is a multivalued function, a careful definition
of $\lambda_n^{-s}$ is required to avoid ambiguities in  (\ref{zeta})
and thence in  (\ref{determinante}). We take it to be $\exp(-
s\log\lambda_n)$ where the cut of the logarithm is chosen to be over the
real positive axis, $0\le{\rm arg}(\lambda_n)<2\pi$, enabling us to rewrite
$\zeta(s)$ in the more convenient form
\begin{equation}
\label{zetacut1}
\zeta(s)=\sum_{{\rm Re}~\lambda_n>0} (\lambda_n)^{-s}+
\exp(-i\pi s)\sum_{{\rm Re}~\lambda_n<0} (-\lambda_n)^{-s}.
\end{equation}
A different cut may alter
the determinant ({\it i.e.}, produce terms that are not proportional to
its intrinsic ambiguity, $\zeta(0)$) only if it intersects the line
${\rm Im}~z=m$ and thereby has crossed an infinite number of
eigenvalues. In that case, instead of (\ref{zetacut1}), one would have
\begin{equation}
\label{zetacut2}
\zeta(s)=\sum_{Re~\lambda_n>0}(\lambda_n)^{-s}+\exp(i\pi s)
\sum_{Re~\lambda_n<0}
(-\lambda_n)^{-s}.
\end{equation}
Eq. (\ref{zetacut2}) has been rewritten by using the same cut as in
(\ref{zetacut1}) in order to compare them; 
we have also dropped contributions  proportional
to $\zeta(0)$. This alternative choice does not affect gauge invariance,
but, as we shall see later, does change the sign of the possible 
parity anomaly terms in $\Gamma[A]$ as was noted in \cite{Rossini}
by more complicated considerations. It
represents the non-perturbative analog of the more familiar sign ambiguity
encountered in defining the perturbative series via, {\it e.g.}, Pauli-Villars
regularization. There, it appears as an explicit dependence
on the sign $M/| M|$ of the regulator mass\footnote{It has
been pointed out
that a larger ambiguity in the perturbative approach can be obtained by using
more than one Pauli-Villars field \cite{PV}.
This unnatural result has a (likewise unnatural) counterpart in  
$\zeta-$function regularization: use the well-known ``product anomaly''
$\det(A B)\ne\det(A)\det(B)$ to bring in definitions that differ
in the number of determinants,  each of whose cuts is  to be
separately fixed.}. We will return to the significance and fixing of the
ambiguity.

Turning now to  gauge invariance in this  framework, it is clear that 
it hinges on  that of the  eigenvalue
spectrum. But small transformations do not affect the $\lambda_n$ at
all, while the large ones merely permute them, as in usual illustrations
of index theorems \cite{pop}. Thus every well-defined symmetric function
of the spectrum, such as $\zeta(s)$ and hence $\Gamma[A]$ is unchanged
and so gauge invariant. This  argument does not rely on the particular
topology of the manifold we are considering, and  it will hold
for finite temperature space-times that are  products of
$S^1$  times a $d-1$ dimensional compact manifold.

To investigate the properties of the determinant more closely, we must
rewrite the "abstract'' $\zeta-$function in terms of the well-established
machinery of the heat-kernel equation. [This task is not completely trivial
because our operator is not positive-definite.] Let us illustrate this 
first in the massless case and then proceed to the massive one. This will
also allows a simpler connection to earlier results.

At $m=0$, the eigenvalues being real, a parity transformation is simply
\begin{equation}
\lambda_n\to-\lambda_n,
\end{equation}
so we can decompose  $\zeta(s)$ into
\begin{equation}
\label{decomposition}
\zeta(s)\equiv\zeta_{PC}(s)+\zeta_{PV}(s),
\end{equation}
parity even and odd parts,
\begin{eqnarray}
&&\zeta_{PC}(s)\equiv\frac{1+\exp(\mp i \pi s)}{2}\sum_{\lambda_n\in
{\rm Spectrum}}( |\lambda_n|)^{-s}\\
&&\nonumber\\
&&\zeta_{PV}(s)\equiv\frac{1-\exp(\mp i\pi s)}{2}\left(\sum_{\lambda_n>0}
(\lambda_n)^{-s}-\sum_{\lambda_n<0} (-\lambda_n)^{-s}\right)
\end{eqnarray}
while the
$\mp$ keeps track of the relevant ambiguity in changing cut. These two
objects can be now easily related to the square ${D\!\!\!\!/}^2$
of the Dirac  operator  (the Laplacian on the spinors) and to the
$\eta-$function of ${D\!\!\!\!/}$. Explicitly we have
\begin{equation}
\label{pippola}
\left\{ \begin{array}{l}
\zeta_{PC}(s)=\displaystyle{\frac{1+\exp(\mp i\pi s)}{2}}\zeta_{{D\!\!\!\!/}^2}
(s/2)\\
\\
\zeta_{PV}(s)=\displaystyle{\frac{1-\exp(\mp i \pi s)}{2}}\eta(s)
\end{array}\right.    .
\end{equation}
Both $\zeta_{{D\!\!\!\!/}^2}(s)$ and $\eta(s)$ are well-defined and
gauge invariant quantities, which admit an explicit  heat-kernel 
representation
\begin{eqnarray}
&&\zeta_{{D\!\!\!\!/}^2}(s)=\frac{1}{\Gamma(s)}\int^\infty_0~dt~t^{s-1}~
{\rm Tr}[\exp(-t{D\!\!\!\!/}^2)],\\
&&\nonumber\\
&&\eta_(s)=\frac{1}{\Gamma({s+1\over 2})}\int^\infty_0~dt~
t^{\frac{s+1}{2}-1}~
{\rm Tr}[{D\!\!\!\!/}\exp(-t{D\!\!\!\!/}^2)].
\end{eqnarray}
Since both functions are analytic\footnote{While 
the regularity of $\zeta_{{D\!\!\!\!/}}^2$ is to  be expected, that 
of $\eta(0)$ is a nontrivial result and we refer the 
interested reader to \cite{Gilkey2}.} at $s=0$, the Dirac determinant
takes the form
\begin{equation}
\label{durament}
\det({D\!\!\!\!/})=\exp\left [-\frac{1}{2}\zeta^\prime_{{D\!\!\!\!/}^2}(0)
\mp\frac{i\pi}{2}\eta (0)\pm\frac{i\pi}{2}\zeta_{{D\!\!\!\!/}^2}(0)\right].
\end{equation}
While the $\mp$ in front of $\eta(0)$ represents a relevant
ambiguity in the definition of the determinant, the $\zeta_{{D\!\!\!\!/}^2}
(0)$ contribution  can be reabsorbed in the first term  by choosing  the
scale parameter$^5$   to be $\mu=-1$.
[Note, in fact, that (\ref{pippola}) implies $\zeta(0)=
\zeta_{{D\!\!\!\!/}^2}(0)$; also $\zeta(0)$  vanishes at odd $d$, being
essentially the conformal anomaly.] In even dimensions  the existence of 
$\gamma_{d+1}$, which anticommutes with the Dirac operator, entails   
the absence of spectral asymmetry and thus the vanishing of $\eta(0)$
so that also the first  $\mp$ is harmless. In  odd dimensions
( no $\gamma_{d+1}$)  no symmetry   prevents
us from having $\eta(0)\ne 0$ and  consequently from having anomalous parity-violating
terms in the effective action whose overall sign is not determined.
Unlike $\zeta_{{D\!\!\!\!/}^2}(0)$, $\eta(0)$ cannot be reabsorbed as
its parity is opposite to that of $\zeta^\prime_{{D\!\!\!\!/}^2}(0)$.
While $\eta(0)$ is a gauge
invariant functional of the field,  it is neither
local nor continuous. It can be explicitly computed with the help of
the Atiyah-Patodi-Singer theorem (see {\it e.g.}, \cite{pop}) and consists
of two parts: a continuous local functional given by the appropriate
dimensional (improved) CS action
plus a highly nonlocal discontinuous  contribution given by
the ``index'' $\pi (N_+-N_-)$. Here $N_+$ is the number of positive
eigenvalues that become negative as $A_\mu$ is continuously deformed to some
reference (background) connection\footnote{ While $B_\mu$ can be taken to be zero for trivial bundles,
the interesting abelian case as we have seen always involves a flux and
hence non trivial ones. In this context 
see, {\it e.g.}, \cite{pop}. Indeed, as shown in \cite{00}
the introduction of the reference 
connection is another way to reach the correct $\bar I_{CS}$.} 
$B_\mu$, and viceversa for $N_-$.  
 Note that  (large) gauge invariance is maintained
through a cancellation between the CS action and the nonlocal index
contribution as advertised earlier. The CS lagrangian is a local 
polynomial of dimension
$d$ in the fields and their derivatives, so it should, in principle,
be removable, unlike the index. If we make
this choice, we obviously lose large gauge invariance: under
transformations of winding number $n$ the determinant is multiplied by a
phase factor $\exp(i\pi n)$. Instead,  parity-invariance is recovered
in spite of the surviving index contribution, because while  the 
index changes sign
under  parity,  it is of the form $i\pi\times$ an integer,
which leaves the determinant unchanged. [The effective action
actually changes by  the acceptable phase $ 2  \pi i n$].

We are now ready to deal with the massive case. Let us first note
that the above massless parity decomposition  still holds formally,
but it has lost its physical meaning  because a parity transformation
here means
\begin{equation}
\lambda_n\to -\lambda^*_n.
\end{equation}
The  eigenvalues have, in fact, become complex since the euclidean Dirac
mass is antihermitian: they are given by $\lambda_n+i m$, where
$\lambda_n$ are those of the massless operator (and $i {/\!\!\!\! D}$ is 
hermitian). In this case, by means
of the Mellin representation of the complex power, we can write,
for the parity even and odd parts,
\begin{eqnarray}
\label{bilba1}
&&\zeta_{PC}(s)=\frac{\exp\left(\mp i
\displaystyle{\pi s\over 2}\right)}{\Gamma(s)}\!
\int^{\infty}_0\!\!\!\!dt~ t^{s-1}\cos\left(m t\mp\frac{\pi s}{2}\right)
\sum_{\lambda_n}\exp(-|\lambda_n| t)\\
&&\nonumber\\
&&\zeta_{PV}(s)=\frac{\exp\left(\mp i
\displaystyle{\pi s\over 2}\right)}{i\Gamma(s)}\!
\int^{\infty}_0\!\!\!\!dt~ t^{s-1}\sin\left(m t\mp\frac{\pi s}{2}\right)\left[
\sum_{\lambda_n>0}\exp(-|\lambda_n| t)-\!\!\!\!
\sum_{\lambda_n<0}\exp(-|\lambda_n| t)
\right].\label{bilba2}
\end{eqnarray}
The kernels in eqs. (\ref{bilba1}) and (\ref{bilba2}) can be again
written in term of the heat-kernels of the { square} of the
massless Dirac operator and of its $\eta-$function. We shall begin by
considering the parity-conserving part $\zeta_{PC}(s)$. The first step
is to find a function $F(s,t)$ such that
\begin{equation}
\label{F(s,t)}
\int^\infty_0 dt\,t^{s-1}\cos\left(m t\mp\frac{\pi s}{2}\right)
\exp(-\lambda t)=\int^\infty_0 dt\, F(s,t)\exp(-\lambda^2 t).
\end{equation}
This integral equation  can be easily solved by interpreting it
as an identity between Laplace transforms. In fact one can immediately
write
\begin{equation}
\label{Luca}
F(s,t)=\frac{1}{2\pi i}\int^{\gamma+i\infty}_{\gamma-i\infty}d\lambda
\exp(\lambda t)\int^\infty_0 dp\,p^{s-1}\cos\left(m p\mp\frac{\pi s}{2}\right)
\exp(-\sqrt{\lambda} p),
\end{equation}
where $\gamma$ is a real constant that exceeds the real part of all the
singularities of the second integral. With the help of $F(s,t)$, the 
$\zeta_{PC}(s)$ takes
the form
\begin{eqnarray}
\zeta_{PC}(2s)&&=\frac{1}{\Gamma(s)}\int^\infty_0\!\!\!\! dt~t^{s-1}~
K_{\mp}(t,s)\sum_{\lambda_n}\exp(-t \lambda_n^2)\nonumber\\
&&=\frac{1}{\Gamma(s)}\int^\infty_0\!\!\!\!dt~t^{s-1}~K_{\mp}(t,s)
{\rm Tr}[\exp(-t{D\!\!\!\!/}^{~2})],
\label{pipa1}
\end{eqnarray}
where
\begin{eqnarray}
\label{Kappa}
&&K_{\mp}(t,s)\equiv\exp(\mp i\pi s)(K^{(1)}(t,s)\pm
2m\sqrt{t}~K^{(2)}(t,s))=\\
&&\exp(\mp i\pi s)\left(\cos(\pi s)
\Phi\left(\frac{1}{2}+s,\frac{1}{2};-m^2 t\right)\pm 2 m\sqrt{t}~
\frac{\Gamma(1+s)}{\Gamma
\left(s+\frac{1}{2}\right)}
\sin(\pi s)\Phi\left(1+s,\frac{3}{2};-m^2 t\right)\right)\nonumber
\end{eqnarray}
and  $\Phi(\alpha,\beta ; z)$ denotes the confluent
hypergeometric function.  Let us now perform the analogous analysis
for the parity-violating contribution $\zeta_{PV}(s)$. This time we
need a function $F(s,t)$  satisfying the integral identity
\begin{equation}
\int^\infty_0 t^{s-1}\sin\left(m t\mp\frac{\pi s}{2}\right)
\exp(-\lambda t)=\int^\infty_0  F(s,t)
\lambda\exp(-\lambda^2 t).
\end{equation}
The explicit form of this $F(s,t)$ can be constructed, as before,
by means of the Laplace transform. In particular we get
\begin{equation}
F(s,t)=\frac{1}{2\pi i}\int^{\gamma+i\infty}_{\gamma-i\infty}
\frac{d\lambda}{\sqrt{\lambda}}
\exp(\lambda t)\int^\infty_0 dp\,p^{s-1}\sin\left(m p\mp\frac{\pi s}{2}\right)
\exp(-\sqrt{\lambda}p),
\end{equation}
with $\gamma$ as in eq. (\ref{Luca}). In terms
of this new kernel, the parity-violating part becomes 
\begin{eqnarray}
\label{pipa2}
\zeta_{PV}(s)&=&\pm\frac{i}{\Gamma\left(\frac{s+1}{2}\right)}\int^\infty_0
\!\!\!\!dt~
t^{\frac{s-1}{2}}~G_{\mp}(t,s)
\sum_{\lambda_n}\lambda_n \exp(-t \lambda_n^2)]\nonumber\\
&=&\pm\frac{i}{\Gamma\left(\frac{s+1}{2}\right)}\int^\infty_0\!\!\!\!dt~
t^{\frac{s-1}{2}}~G_{\mp}(t,s)
{\rm Tr}[{D\!\!\!\!/}\exp(-t{D\!\!\!\!/}^2)],
\end{eqnarray}
with
\begin{eqnarray}
\label{Kappa2}
&&G_{\mp}(t,s)\equiv \exp\left(\mp i\frac{ \pi s}{2}\right)\sin\left(
\frac{\pi s}{2}\right)(G^{(1)}(t,s)\mp 2 m\sqrt{t}~G^{(2)}(t,s)) =\\
&&\exp\left(\mp i\frac{ \pi s}{2}\right)\sin\left(
\frac{\pi s}{2}\right)\left(\Phi\left(\frac{s}{2},\frac{1}{2};-m^2 t
\right)\mp 2 m\sqrt{t}~
\frac{\Gamma\left(1-\displaystyle{\frac{s}{2}}\right)}{\Gamma
\left(\displaystyle{\frac{1-s}{2}}\right)}
\Phi\left(\frac{1+s}{2},\frac{3}{2};-m^2 t\right)\right).\nonumber
\end{eqnarray}
Eqs. (\ref{pipa1}) and (\ref{pipa2}) are the promised ``spectral'' 
representations for the $\zeta$ and $\eta$ functions, and in particular 
the weights $G_{\mp}$ and $ K_{\mp}$ encode all the information
about the mass dependence of our determinant. [Actually they
contain more, because they hold for all $s$ and not
only at $s=0$.] Therefore they can be used to investigate the properties
of the effective action in different mass limits. In the next section
 we shall use them to derive  expansions of the effective
action for small and large masses. With their help, one can also show that
the general considerations developed in the massless case extend unchanged to
the massive one.

\section{ Large and Small Mass Expansions}
\setcounter{equation}{0}
The parity-conserving part of the effective action  is given by
\begin{eqnarray}
\label{PCA}
\Gamma_{PC}[A]=\left.\frac{d}{ds}\zeta_{PC}(s)\right|_{s=0}\!\!\!\!\!\!\!&=&\!\!
\frac{1}{2}\left.\frac{d}{ds}\zeta_{PC}(2s)\right|_{s=0}\!\!\!\!\!\!\!=\!\!
\frac{1}{2}\frac{d}{ds}\left[\displaystyle{\frac{1}{\Gamma(s)}}
\int^\infty_0\!\!\!dt~t^{s-1} K_{\mp}(t,0)
{\rm Tr}[\exp(-t{D\!\!\!\!/}^{~2})]\right]_{s=0}\nonumber\\
&&\nonumber\\
&&+\lim_{s\to 0}\displaystyle{\frac{1}{\Gamma(s)}}
\int^\infty_0 ~dt~t^{s-1}\frac{d K_{\mp}}{ds}(t,s)
{\rm Tr}[\exp(-t{D\!\!\!\!/}^{~2})].
\end{eqnarray}
The limit $s\to 0$ in  (\ref{PCA}) is a delicate point, detailed in appendix A. 
The final result is 
\begin{eqnarray}
\label{pqpq1}
\Gamma_{PC}[A]&=&\!\!\frac{1}{2}\left.\frac{d}{ds}
\zeta_{{D\!\!\!\!/}^{~2}+m^2} (s)
\right|_{s=0}\!\!\!\!\mp\frac{\pi i}{2}\zeta_{{D\!\!\!\!/}^{~2}+m^2}(0)+
\nonumber\\
&&+\ \ \left\{\begin{array}{l}
\displaystyle{\Gamma^{Odd}_{Loc}[A]=\pm {\rm sign}(m)\sqrt{\pi}
\sum_{k=0}^{(d-1)/2} 
\frac{(-2)^{k} }{(2k+1)!!}(m^2)^{k+1/2} H_{d-1-2 k},}\\
\\
\displaystyle{\Gamma^{Even}_{Loc}[A]=-\sum_{k=1}^{d/2} 
\left [\sum_{j=1}^{k} \frac{(-2)^{j-1}}{j}
\frac{1}{(2j-1)!!(k-j)!}\right ] (im)^{2k} H_{d-2 k} },\\
\end{array}\right.
\end{eqnarray}
where $H_n$ are the Seeley--deWitt  \cite{00}
coefficients for the massless 
Laplacian on the spinor: ${\rm Tr}[\exp(-t{D\!\!\!\!/}^{~2})]=
\displaystyle{\sum_{n=0}^{\infty}} H_n~t^{(n-d)/2}$. That the 
non-local part of the parity-conserving action 
($1/2 \zeta^\prime_{{D\!\!\!\!/}^{~2}+m^2} (0)$) is governed by 
the massive Laplacian might be expected, but, surprisingly, we 
have extra dimension-dependent local contributions coming from 
the $s-$derivative of the kernel $K_{\pm}(s,t)$. Note 
that in odd  dimensions, in contrast to the even ones, their sign 
depends on the choice of cut. This phenomenon will become more 
relevant for the parity-violating part.

The analysis of the parity-violating effective action is substantially 
easier due to the absence of singular contributions as $s\to 0$; one obtains
\begin{equation}
\Gamma_{PV}[A]=\left. \frac{d}{ds} \zeta_{PV}(s)\right |_{s=0}=
\lim_{s\to 0}\pm\frac{i}{\Gamma\left(\frac{s+1}{2}\right)}
\int^\infty_0\!\!\!\!dt~
t^{\frac{s-1}{2}}~\frac{d G_{\mp}(t,s)}{d s}
{\rm Tr}[{D\!\!\!\!/}\exp(-t{D\!\!\!\!/}^{~2})].
\end{equation}
The derivative of the kernel $G_{\mp}(t,s)$ at $s=0$ can be explicitly 
computed and gives
\begin{eqnarray}
\left.\frac{d G_{\mp}(t,s)}{d s}\right |_{s=0}&=&
\frac{\pi}{2}\left[\Phi\left(0,\frac{1}{2},- m^2 t\right)\mp \frac{2}
{\sqrt{\pi}}m\sqrt{t}
\Phi\left(\frac{1}{2},\frac{3}{2},-m^2 t\right)\right]=\nonumber\\
&=&\frac{\pi}{2}\left(1\mp \frac{2{\rm sign}(m)}
{\sqrt{\pi}}\int^{m\sqrt{ t}}_0 \exp(-z^2) d z\right).
\end{eqnarray}
Thus  the parity-violating part of the action turns out to be
(in odd d, where it exists)
\begin{eqnarray}
&&\Gamma_{PV}[A]=\pm\frac{i\pi}{2}\eta(0)-i{\rm sign}(m)
\int^{\infty}_0\frac{dt}{\sqrt{t}}{\rm Tr}[{D\!\!\!\!/}\exp(-t
{D\!\!\!\!/}^{~2})]
\int^{|m|\sqrt{t}}_0 dz \exp(-z^2)=\nonumber\\
&&(\pm 1-{\rm sign}(m))
\frac{i\pi}{2}\eta(0)+i{\rm sign}(m)
\int^{\infty}_0\frac{dt}{\sqrt{t}}{\rm Tr}[{D\!\!\!\!/}\exp(-t
{D\!\!\!\!/}^{~2})]\int_{|m|\sqrt{t}}^\infty dz \exp(-z^2).
\label{popo}
\end{eqnarray}
In $d=3$ a similar representation for $\Gamma_{PV}[A]$
was given in \cite{Leutwyler}. There, the cut giving the
plus sign  was implicitly chosen. As we shall see below, this
corresponds to requiring  ``decoupling'', {\it i.e.} vanishing of 
$\Gamma$, as the fermion mass goes
to $+\infty$. Note, again, that the sign in front of the parity anomaly
is entirely dependent on the choice of branch. As is clear from its
representation our $\Gamma_{PV}$ differ from the odd-mass part 
$\tilde{\Gamma}_{PV}$ of
$\Gamma$,
\begin{equation}
\tilde \Gamma_{PV}[A]\equiv\frac{1}{2}\left (\Gamma[A,m]-\Gamma[A,-m]\right)
\label{xxx}
\end{equation}
as $\Gamma_{PV}$ has even mass parts as well  (and $\Gamma_{PC}$ odd ones).
This $\tilde \Gamma_{PV}$ clearly cannot detect the intrinsic anomaly (the
one at $m\equiv 0$); as a result the possibility of decoupling in the infinite
mass limit is not manifest. [Also in a nonflat background geometry or higher
dimensions, the  above definition actually contains parity-conserving terms.]

Both  (\ref{pqpq1}) and (\ref{popo}) can be used as  starting
points for a mass expansion of the theory. Let us first consider
the small mass limit: in the parity conserving case we simply have to
Taylor-expand  $\zeta_{{D\!\!\!\!/}^{~2}+m^2}(s)$ in power of mass.
For odd dimension
\begin{equation}
\Gamma^{odd}_{PC}[A]=\frac{1}{2}{d\over ds}
\zeta_{{D\!\!\!\!/}^{~2}}(0)+\sum_{k=1}^{\infty} (i m)^{2 k}
\frac{\zeta_{{D\!\!\!\!/}^{~2}}(k)}{2k}+\Gamma^{odd}_{Loc}[A],
\label{corno1}
\end{equation}
where $\Gamma^{odd}_{Loc}[A]$  is specified in eq. (\ref{pqpq1}). The appearance
of the even power can be understood as a consequence of the behavior of the Dirac
mass term under parity. Instead, the local contributions ($\Gamma^{odd}_{Loc}[A]$),
proportional to $m^{2k+1}$,
originate from a compensation between vanishing and
divergent terms as $s$ goes to zero. The even dimensional case is more 
delicate, due to the fact that $\zeta_{{D\!\!\!\!/}^{~2}}(s)$ has in general
simple poles for $n=1,2,...,{d}/{2}$. The final result can be presented in the
form
\begin{eqnarray}
\Gamma_{PC}[A]&&=\frac{1}{2}{d\over ds}
\zeta_{{D\!\!\!\!/}^{~2}}(0)+\sum_{k=\frac{d}{2}+1}^{\infty} (i m)^{2 k}
\frac{\zeta_{{D\!\!\!\!/}^{~2}}(k)}{2k}\nonumber\\
&&+\sum_{k=1}^{\frac{d}{2}}\frac{1}{2}\frac{d}{ds}[\frac{1}{\Gamma(s)}
\int_{0}^{\infty}dt\,t^{s+n-1}{\rm Tr}(\exp-t{D\!\!\!\!/}^{~2})]
+\Gamma^{even}_{Loc}[A].
\label{corno2}
\end{eqnarray}
Analogously, Taylor-expanding the parity violating part, we obtain  
\begin{equation}
\label{potta2}
\Gamma_{PV}[A]=
	\pm i\frac{\pi}{2} \eta(0)-
i \sum_{k=0}^\infty  (-1)^k\frac{m^{2 k+1}}{2k+1}
\eta (2 k+1).
\end{equation}

Note the presence of the intrinsic parity anomaly term $\displaystyle{
\pm i\frac{\pi}{2} \eta(0)}$: it is the only one proportional to an even,
$m^0$, power of the mass.    We have already
stated that it  contains the CS action, but this does 
$\underline{\rm not}$ mean that there are no other CS
contributions  hidden in the rest of the series! The large mass
analysis below  and the examples in sec. 5 will indicate that they are
actually present. Furthermore their coefficients are  obviously mass- and
consequently temperature-dependent ( the
mass can appear only through a dimensionless combination such as $\beta m$,
though other combination are possible if there are other relevant  scales in
the problem, {\it e.g.}, the volume of the manifold). On the other hand,
 gauge invariance is entirely unaffected by this: each term in the
series is manifestly gauge invariant, since $\eta(s)$ is. 
 
The large mass limit is a more delicate issue,  corresponding to
an asymptotic expansion of the action. In the case of $\Gamma_{PC}$, a
simple application of Watson's lemma\footnote{  
It essentially states \cite{pippi} that an asymptotic expansion in $m$ of 
integrals like $\displaystyle{\int^{\infty}_{-\infty} dt e^{-t m^2} f(t)}$ 
can be obtained by integrating the asymptotic expansion of $f(t)$ term by 
term.}  gives in the odd-dimensional case
\begin{equation}
\Gamma_{PC}[A]=\sqrt{\pi}(-1\pm{\rm sign}(m))\sum_{n=0}^{(d-1)/2}
\frac{(-2)^n}{(2n+1)!!}
(m^2)^{n+\frac{1}{2}} H_{d-1-2n}
+\frac{1}{2}\sum_{n=\frac{d+1}{2}}^\infty
\frac{\Gamma\left(n-\displaystyle{\frac{d}{2}}\right)}{(m^2)^{n-d/2}}H_{2n},
\label{mpclarge1}
\end{equation}
while in even dimensions we have
\begin{eqnarray}
\Gamma_{PC}[A]&=&
\displaystyle{\sum_{n=d+1/2}^\infty}
\frac{\Gamma\left(n-\displaystyle{\frac{d}{2}}\right)}{(m^2)^{n-d/2}}H_{2n}
+\frac{1}{2}
\displaystyle{\sum_{n=0}^{d/2-1}}
\frac{(-m^2)^{d/2-n}}{\Gamma\left(\displaystyle{\frac{d}{2}}+1-n\right)}
\left[\sum_{k=1}^{d/2-n}\frac{1}{k}-\log\frac{m^2}{\mu^2}\right]H_{2n}
\nonumber\\
&&-\frac{1}{2}\log\frac{m^2}{\mu^2}H_{d}-\sum_{n=0}^{d/2-1}
(-m^2)^{\frac{d}{2}-n}\left[\sum_{k=1}^{\frac{d}{2}-n}
\frac{1}{(2k-1)!!}\frac{(-2)^{k-1}}{k(\frac{d}{2}-n-1)!}\right]H_{2n}. 
\label{mpclarge2}
\end{eqnarray}
Essentially, to obtain (\ref{mpclarge1}) and (\ref{mpclarge2}), one expands
the kernel ${\rm Tr}[\exp(-t{D\!\!\!\!/}^{~2})]$ for small $t$ and
integrates term by term.  The asymptotic nature of this series means that
terms exponentially small in the mass, {\it i.e.} $O(e^{-\beta m})$,  
cannot be seen.
This can have quite dramatic consequences, as we will show through explicit 
examples in sec. 5.  Nevertheless the expansion is both large and small 
gauge invariant order by order.
In  (\ref{mpclarge1}) and (\ref{mpclarge2}) we have inserted the explicit 
form of the local terms:  we mention first that in odd dimensions 
the divergent contributions (in the large mass limit) are non-vanishing only 
when gravity (through the geometry of the manifold) is involved. This can 
be inferred from  the structure of the heat-kernel 
coefficients: for example in three dimensions $H_1$ and $H_2$ correspond to
the cosmological term and to the Einstein action respectively. In general 
their coefficient is strongly dependent on the cut. For positive mass,
the branch chosen in (\ref{zetacut1}) gives zero ({\it i.e.}, the
fermion decouples),  while the one in (\ref{zetacut2}) would give
a limit value of $2$ (no decoupling). For negative mass, the reverse situation
occurs with coefficient $(-2,0)$. This  shows vividly that the choice of the
cut is not just a matter of convention, but  affects physical
predictions. It is interesting to notice that in $d=3$ 
the first non-trivial correction to the infinite mass limit 
(the $H_4$ coefficient) is a Maxwell 
$(F^2_{\mu\nu})$ term, with  coefficient $\displaystyle{\frac{1}{48\pi}
\frac{1}{|m|}}$, in agreement with earlier calculations 
\cite{Red,Leutwyler,pipponie}.
In the even-dimensional case the expansion is independent of the cut, 
as one would expect, and also involves  logarithmic dependence on the mass, 
due to the non-vanishing of the trace anomaly there.

The analysis of the
behavior of the parity-violating part is more intricate, and
a straightforward application of  Watson's  lemma is not possible.
However, looking at  (\ref{popo}), one realizes that, for large
mass, only small $t$ can contribute. The large $t$ behavior is, in
fact, suppressed by the vanishing of the error function. Thus we
can again expand the kernel and integrate term by term. This time we use 
the heat-kernel expansion $\displaystyle{
{\rm Tr}[{D\!\!\!\!/}\exp(-t{D\!\!\!\!/}^{~2})]=
\sum_{n=0}^{\infty}P_{n}~t^{(n-d-1)/2}}
$, where $P_n$ are different from zero only for odd $n$. 
\noindent
We therefore obtain
\begin{equation}
\label{potta3}
\Gamma_{PV}(A)=(-1\pm{\rm sign}(m))\frac{i\pi}{2}\eta(0)
+i\sum_{n=0}^\infty 
\frac{P_{2n+1}}{(m^2)^{n-(d-1)/2}}
\frac{\Gamma\left(n-\displaystyle{\frac{d}{2}}+1\right)}{2n+1-d}.
\end{equation}
Let us stress again the asymptotic nature of this
series. The  (local and invariant functionals of the gauge fields and 
of the  geometry) $P_{2n+1}$ coefficients  differ from zero only for 
$n>(d-1)/2$. 
In the limit of infinite mass, the only possible surviving term is 
therefore proportional to the gauge invariant $\eta(0)$, but different
coefficients are possible, in complete analogy with the parity
conserving sector: $(2,0)$ for large positive mass and the cut as
in (\ref{zetacut1}); $(0,-2)$ for large negative mass and the cut
as in (\ref{zetacut2}). Thus,
given a sign of the fermion mass, the branch can be always chosen so that the
fermion completely decouples (or not !) in the infinite mass 
limit.  This double pair of possibilities completely
mimics the analogous perturbative result in the
presence of one Pauli-Villars regulator. There
the final asymptotic result would have been
\begin{equation}
\Gamma_{PV} [A]\simeq [{\rm sign}(m)+{\rm sign}(M)] I_{CS},
\end{equation}
where $m$ is the fermion mass, while $M$ is the mass of the regulator.
For $m$ positive, we have $(2,0)$ as $M\to(+\infty,-\infty)$, for
$m$ negative we have instead $(0,-2)$ as  $M\to(+\infty,-\infty)$.
The absence of the index in the perturbative result
implies the loss of manifest gauge invariance for  finite
masses since  $I_{CS}$ has no counterpart to restore it (nor
does it acquire the required boundary terms needed to make it 
well-defined).

\medskip
\section{Explicit Gauge field Examples}

\setcounter{equation}{0}

For  concrete illustrations of how the perturbative paradox is
circumvented, let us now consider some explicit examples  of
actions and large gauge transformations. We start by reviewing,
according to \cite{us},
the $(0+1)-$dimensional toy model of ref. \cite{Dunne}. It consists
of $N$ fermions on a circle of a radius $\beta$ interacting with a 
$U(1)$-field through
the Lagrangian
\begin{equation}
\label{bibi}
{\cal L}=\sum_{i=1}^N\bar\psi_i(t)\left(i\frac{d}{d t}+ A(t)+ i m
\right )\psi_i(t).
\end{equation}
 The   large transformations are taken to be
\begin{equation}
U(t)=\exp(i f(t)),\ \ \ \ \ \ \ \ {\rm where} ~~f(\beta)-f(0)=2\pi n.
\end{equation} 
The integer $n$ is the winding number of the map $U(t)$, {\it i.e.},
$2\pi in=\displaystyle{\int}^\beta_0 ~dt  U(t)^{-1} U^\prime (t)$.
The analog of parity in three dimensions is here charge conjugation
$A\to -A$; while  massless fermions are  invariant,  massive ones
violate this symmetry. [Had a bare CS term, here $ k A(t)$, been present
in (\ref{bibi}), invariance of the path-integral under large 
transformations would require that $k$ in (\ref{bibi}) be quantized,
entirely  as in $D=3$.] 
 
\noindent
The eigenvalue problem corresponding to the $(0+1)-$dimensional Dirac
operator can be exactly solved, 
\begin{equation}
\lambda_n= \frac{2\pi}{\beta}\left (n-\frac{1}{2}\right)+\frac{2\pi}{\beta} a+ i m
\ \ \ \ \ \ n\in Z\!\!\!Z,
\end{equation}
where $a$ is the average  of $A$: $\displaystyle{ a=\frac{1}{2\pi}
\int^\beta_0 A(t) dt}$. The $\zeta-$function can be computed in  
closed form in terms of the Hurwitz function \cite{Tavole} $\zeta_H(s,q)$ 
\begin{equation}
\zeta(s)=N\left(\frac{\beta}{2\pi}\right)^s
\left[\zeta_H \left(a+\frac{1}{2}+i\frac{\beta m}{2\pi},s\right)
+\exp(\mp i\pi s) \zeta_H \left(\frac{1}{2}-a-i\frac{\beta m}{2\pi},
s\right). \right]
\end{equation}
Throughout, the $\mp$ keeps track of the relevant
ambiguity in choice of cut. The determinant  is now easily evaluated
directly from its definition,
\begin{eqnarray}
\label{pipo}
&&\exp\left(-\Gamma(A)\right)=
\det\left(i\frac{d}{d t}+ A(t)+ i m\right )=\exp(-\zeta^\prime (0))=
\left[2\left (\cosh\left(\frac{\beta m}{2}\right)\cos{\pi a}\right.
\right.\nonumber\\
&&\left.\left.
-i \sinh\left(\frac{\beta m}{2}\right)\sin{\pi a}\right )
\exp\left(\pm i\pi a\mp\frac{\beta m}{2}\right)\right]^N\equiv
\left [1+e^{\pm (2\pi i a-\beta m)}\right ]^N.
\end{eqnarray}
Note that  this action depends on $a$ only via the $S^1$ holonomy
$\exp(2\pi i a)$ and thus is manifestly gauge invariant under a
{\it large} transformation, $a\to a+1$, for either cut
and for any $N$, even or odd. In the middle term this occurs through
a sign cancellation between the separate factors.
Though the value of final expression in (\ref{pipo})
seems to  depend completely on the choice of cut, the intermediate equality
makes it clear that only the charge conjugation anomalous contribution is affected,
in agreement with the general results of sec. 3. Notice also the necessary
presence of an ``intrinsic"( {\it i.e.} even present at $m=0$) charge 
conjugation anomaly
${\rm Im}\Gamma_{CV}[A]=iN(a-[a])$, where $[a]$ denotes the integer part
of $a$. This is what allows us to preserve large gauge invariance
independently of $N$. This result also clearly exhibits what was
claimed on general grounds in sec. 3~ for the parity anomalous contribution,
namely the $\eta(0)$: only the combination of the continuous part, given by
the CS action $a$ and the discontinuous contribution coming from the index $[a]$
is gauge invariant. Had we opted instead (as in \cite{Dunne}) for the
$(0+1)$ equivalent of the more usual 
 $C-$preserving regularization, the $\exp(i N\pi a)$ factor 
in (\ref{pipo}) would have been missing and only even $N$ would have kept 
invariance, just as in $(2+1)$.

Dimension $(0+1)$  is also a good laboratory for testing  the mass
expansions discussed in sec. 4 and in particular that for  large mass.
If we apply the one-dimensional analog of parity conserving/violating
expansions (\ref{mpclarge1}) and (\ref{potta3}) (or directly from
(\ref{pipo})), we obtain 
\begin{equation}
\label{fff}
\Gamma[A]\simeq (0,-2)\left[i\pi
(a-[a])-\frac{\beta m}{2}\right],\ \ \ \ \  m>0
\end{equation}
\begin{equation}
\label{fff1}
\Gamma[A]\simeq(2,0)\left[i\pi
(a-[a])-\frac{\beta m}{2}\right],\ \ \ \  \ m<0
\end{equation}
 This is  a concrete 
realization of what was stated  at the end of sec. 4.  Let us
also notice that  all the $1/m$ corrections are identically zero. 
One can understand this result from two different points of view.  Firstly,
beyond the terms shown in (\ref{fff}) and (\ref{fff1}), all the others  are exponentially
small in the mass and thus they cannot affect the asymptotic series. Secondly,
the one-dimensional Dirac operator coupled to a gauge field is always
locally gauge-equivalent to the free one (since locally $A(t)=\partial_t
B(t)$). This means that the  local
coefficient of its heat-kernel expansion must be trivial, and
dramatically shows how much information can be lost in a
large mass expansion, even though the final result is gauge invariant.
In other words, when topological degrees of freedom such as $a$ are
involved, an expansion in the local coefficients of the heat-kernel
can rarely retain the complete dynamics of the theory.

Though very instructive because of its soluble nature, one might wonder
if the mechanism realized in the toy $(0+1)$ model  is shared by its
3-dimensional  counterpart, where a complete  solution of the theory is
not at our disposal. A more realistic example in this direction is to
consider a purely magnetic configuration with flux $\Phi(B)=2\pi n$
in $d=3$. It is an easy exercise to show that the most general potential, 
up to a gauge transformation, generating such a field is 
\begin{equation}
\label{puppa}
A_\mu \equiv\left(\frac{2\pi}{\beta} a, {\bf A}({\bf x})
\right),
\end{equation}
where $a$ is a flat $S^1$ connection  and the $2-$potential ${\bf A}$
is static, living on the two-dimensional Riemann manifold $\Sigma^2$. 
The  large transformations are associated to the  $S^1$ map $a\to a+1$,
 as in the $(0+1)-$dimensional case. [For  $\Sigma$ of genus
greater than 0, large transformations corresponding to the non-trivial
cycles of $\Sigma$ can be also constructed, but we will not discuss
them. Here we will only be interested in the ones relevant in the finite
temperature regime.]

We  now proceed to compute the partition function for a  single 
Dirac fermion in the background (\ref{puppa}). Since the latter is
time-independent, we can decouple $t$  by looking for
eigenvectors of the form 
\begin{equation}
\hat\psi(t,x,y)=\exp\left[-\frac{2\pi}{\beta}\left(n+\frac{1}{2}\right) t 
\right ]\psi(x,y).
\end{equation}
[In finite temperature field theory the integer factor $n$ in the phase is
usually known as the Matsubara frequency.]  The $1/2$ factor 
takes care of the antiperiodic boundary conditions for the fermion.
The eigenvalue problem for the $d=3$ operator $i /\!\!\!\! {\cal D}$
thus reduces to an infinite series of effective
two-dimensional ones parameterized by $n$,
\begin{equation}
\label{pup}
i /\!\!\!\! {\cal D}\psi=i /\!\!\!\!\hat D\psi+
\frac{2\pi}{\beta}\left (\alpha_0+n+\frac{1}{2}\right )\gamma^0\psi=
(\lambda-i~m)\psi\equiv\hat \lambda\psi
\end{equation}
Here $i /\!\!\!\!\hat D$ is the massless Dirac operator on the 
two-dimensional manifold $\Sigma$. 
The key observation is that the spectrum of  
$/\!\!\!\!\cal{ D}$
can be reconstructed once that of  $/\!\!\!\!\hat{ D}$ is known.
In fact let 
\begin{equation}
\phi(x,y)=\left (\phi_1(x,y)\ , \ \phi_2(x,y)\right )
\end{equation}
be a ($2-$component) eigenvector of $/\!\!\!\!\hat{ D}$, with eigenvalue 
$\mu\ne 0$. Then the vectors
\begin{equation}
\psi_{\pm}(x,y)=\left ( 
\phi_1(x,y), C_{\pm}\phi_2(x,y) \right ),
{C_{\pm}=-\frac{2\pi}{\beta\mu}
\left(\alpha_0+n+\frac{1}{2}\right)\pm
\sqrt{\frac{4\pi^2}{\beta^2\mu^2}\left (a+n+\frac{1}{2}\right)^2
+1} }
\end{equation}
 are eigenvectors of $/\!\!\!\!{\cal D}$ with
eigenvalues
\begin{equation}
\lambda_{(\pm)}=i~m\pm\sqrt{\frac{2\pi}{\beta}\left 
(a+n+\frac{1}{2}\right)^2+\mu^2}.
\end{equation}
As might be expected from the $\gamma_0$ in (\ref{pup}), each
non-vanishing eigenvalue of $/\!\!\!\!\hat D$ generates
two eigenvalues of $/\!\!\!\!{\cal D}$ of equal multiplicity.
This  symmetrical behavior suggests that they will not produce
a spectral asymmetry and thus play no role in the clash between
invariances under large and parity transformation. In fact, by using
the representation (\ref{bilba2}) for  $\zeta_{PV}(s)$, it is
immediate to see that their contribution there vanishes.

We come now to  ${\rm Ker}/\!\!\!\!\hat D$.
The Atiyah-Singer theorem tells us that it is spanned by $\nu_+$ 
spinors $\phi^0_+(x,y)$ with positive chirality and $\nu_-$
spinors $\phi^0_-(x,y)$ with  negative chirality, where $\nu_+-
\nu_-=n$ is the flux of the ${\bf A}$. ($\gamma^0\phi^0_\pm(x,y)
=\pm \phi^0_\pm(x,y)$.) Both $\phi^0_+(x,y)$ and $\phi^0_-(x,y)$
are eigenvectors of $/\!\!\!\!{\cal D}$ as well , but with 
eigenvalues
\begin{equation}
\lambda^0_{(\pm)}=i m \pm\frac{2\pi}{\beta}\left(n+a
+\frac{1}{2}\right).
\end{equation}
The chiral asymmetry of the ${\rm Ker}~/\!\!\!\!{\hat D}$ is inherited 
by the spectrum of $/\!\!\!\!{\cal D}$: in fact $\lambda^0_-$ 
and $\lambda^0_+$ have  different degeneracy. This, as we shall
see, will give rise to a non-vanishing  anomalous parity 
contribution.

The ensuing $\zeta-$function is\footnote{Having already noticed that
no asymmetry is entailed by the eigenvalues $\lambda_{(\pm)}$, we have
written $\sum (\lambda_{(+)}\lambda_{(-)})^{-s}$ instead of
$\sum (\lambda_{(+)})^{-s}+\sum (\lambda_{(-)})^{-s}$. 	In fact, in
absence of spectral asymmetry, these two quantities coincide up to
local terms. The difference, proportional to the volume of
$\Sigma$ in this case, can be evaluated with the help of the spectral
representation given in sec. 3.}
\begin{eqnarray}
\label{coppola1}
\zeta(s)&\!\!=\!\!&
\nu_+ \zeta_H\left(\frac{1}{2}+a+i\frac{\beta m}{2\pi},s\right )+
\exp\left(-i\pi s\right )\nu_+ \zeta_H\left(\frac{1}{2}-a-
i\frac{\beta m}{2\pi},s\right )+\!\!\!\!\!\!\\
&&\nu_- \zeta_H\left(\frac{1}{2}-a+i\frac{\beta m}{2\pi},s\right )+
\exp\left(-i\pi s\right)\nu_-\zeta_H\left(\frac{1}{2}+a-i
\frac{\beta m}{2\pi},s\right )
+\nonumber\\
&&\sum_{n,\mu_k}\  
\left [m^2+\frac{4\pi^2}{\beta^2}\left 
(a+n+\frac{1}{2}\right)^2+\mu_k^2\right] ^{-s},\nonumber
\end{eqnarray}
where the discrete sum runs over $n\in Z\!\!\!Z$ and $\mu_k\in [ {\rm Spec}({/\!\!\!\!
{\hat D}})-{\rm Ker}({/\!\!\!\!{\hat D}})]$. 
Let us denote the sum term by the symbol $\zeta_{/\!\!\!\!{\hat D}^2}(s)$
even though that identification is not entirely correct. The determinant can be
then computed and we obtain\footnote{After the derivation in \cite{us} of the general form (\ref{colbacco}) for the
effective action, its odd-mass part $\tilde\Gamma_{PV}$ (\ref{xxx}) 
(rather than the true $\Gamma_{PV}$ itself)  was
recalculated in [30a] in a different way. The result there, which was its main
content, was incorrect. Upon private explanation of  their mistake to the
authors,  a second, corrected, version $[30b]$ properly 
acknowledged our corrections. However,  that acknowledgment
did not survive in the published version $[30c]$, nor in its erratum $[30d]$
stating the true date of  the revised version
$[30c]$. }$^,$\footnote{A recent computation,
\cite{Fosco}, of $\tilde\Gamma_{PV}$ agrees with that implied by (\ref{colbacco})}
\begin{equation}
\label{colbacco}
\exp(-\Gamma(A))=
\left [\exp(-\beta m+2\pi i a)+1\right]^{\nu_+}
\left [\exp(-\beta m-2\pi i a)+1\right]^{\nu_-}
\exp(-\zeta^\prime_{/\!\!\!\!{\hat D}^2}(0)).
\end{equation}
From eq.(\ref{colbacco}) it is manifest that the determinant splits in the
product of two $(0+1)-$dimensional contributions and a reduced expression
depending on ${\bf A}$, $\Sigma$ and the flat connection $a$. Amusingly,
one can go further and partially compute $-\zeta^\prime_{/\!\!\!\!{\hat D}^2}
(0)$, namely  perform the sum over $n$. To this end, one first defines a
Mellin representation of the complex power and then Poisson-resums the
series in $n$ [see appendix B]. In this way, we end up with a 
series for $\zeta_{/\!\!\!\!{\hat D}^2}(s)$ that is analytic at $s=0$ and 
whose derivative at $s=0$ leads to
\begin{eqnarray}
\label{coppola3}
\exp(-\zeta_{/\!\!\!\!{\hat D}^2}(s))&=&
\left |\prod_{\mu_k}\left(1+\exp\left(-\beta\sqrt{\mu^2_k+m^2}
+2\pi  i a \right)\right)\right |^2\ 
\exp\left[2\pi ~{\cal F}
- (\nu_+ +\nu_-) m \beta\right ]\nonumber\\
{\cal F}&=&\zeta_{\frac{\beta^2}{4\pi^2}({/\!\!\!\!{\hat D}}^2+m^2)}(-1/2)
\end{eqnarray}
That the above infinite product is convergent follows immediately
from the  estimate (\ref{stima}). We have thus provided the explicit general 
form (\ref{colbacco}, \ref{coppola3}), for the complete effective action in the 
background (\ref{puppa}). It is a trivial exercise to compute in particular 
its parity-violating part (under $a\rightarrow -a$). The term governed by 
$\zeta_{/\!\!\!\!{\hat D}^2}(s)$ is unaffected, so we obtain
\begin{equation}
\Gamma_{PV}[A]=(\nu_+-\nu_-)\left[\arctan\left(\tan ({\beta m\over 2})
\tan (\pi(a-[a]))\right)\pm \pi(a-[a])\right].
\end{equation}
The above equation exhibits the remarkable property of $\Gamma_{PV}[A]$ 
that it factorizes into a part dependent only on the holonomy $a$ 
times one that involves, through $(\nu_+-\nu_-)$, the flux $\Phi$
on  $\Sigma^2$ since as we saw
$(\nu_+ -\nu_-)= \Phi /2\pi$. This is both in accord with our initial ``Fourier'' representation 
as well as a general consequence of the index theorem on product manifolds 
(for details see \cite{Gilkey2} p. 288). [ We have written the 
redundant combination $(a-[a])$ rather than $a$ in the argument of the $\tan$
above to emphasize its fundamental role.]

A simple but interesting special case of 
(\ref{puppa}) where the eigenvalues $\mu_k$ are known explicitly is the
instanton on the flat unit torus: $A_i=-\pi n \epsilon_{ij}x^j$.
Here $\mu^2_k=4\pi |n k|$ with $2n$ degeneracy, while $2\pi\zeta_{\frac{\beta^2
}{4\pi^2}({/\!\!\!\!{\hat D}}^2+m^2)}(-1/2)=n\left({4\pi n}
\right)^{1/2}\beta~\zeta_H\left (-1/2,\frac{m^2}{2\pi n}\right)
-(\nu_+ +\nu_-) m \beta$. Substituting into (\ref{coppola3})
we obtain
\begin{eqnarray}
&&\exp(-\Gamma(A))=\\
&&\left |\prod_{k=1}^\infty\left(1+\exp\left(-\beta\sqrt{4\pi |n k|+m^2}
+2\pi  i a \right)\right)\right |^{4n}\!\!\!\! 
\exp\left[n\left({4\pi n}
\right)^{1/2}\beta~\zeta_H\left (-1/2,\frac{m^2}{2\pi n}\right)
 - 2 m n\beta\right ].\nonumber
\end{eqnarray}

\noindent 
There are a number of other informative general properties to be drawn 
manifest and its structure is consistent with (\ref{coppola0}). 
Second, it is clear that a perturbative (i.e., in powers of $a$)
expansion of (\ref{colbacco}) and (\ref{coppola3}) loses
periodicity in $a$ and hence does not see large invariance order by 
order. For example the Chern-Simons term  ( $I_{CS}=\pi a n$) has a 
coefficient $1-\tanh\left(\frac{\beta m}{2}\right )$. 
The usually quoted coefficient omits the $``1''$ that represents the 
intrinsic parity-anomaly price of our gauge-invariant regularization and hence
persists at $m=0$. As we saw in sec.3 there is actually an ambiguity in its 
sign (reflecting
a choice of cut), also present in other regularizations,
for example through the factor $\lim_{M\to\pm\infty} {\rm sign}(M)$ in 
Pauli-Villars, even at perturbative Feynman diagram level. 
As we discussed, with our intrinsic parity-violating gauge-preserving 
choice the ambiguity is physically reflected in the degree of decoupling of 
a heavy fermion.

\section{Conclusions}

\setcounter{equation}{0}

After first deriving a generic form for the abelian gauge field 
effective action in 
arbitrary dimensions purely  on gauge invariance grounds, we were
able to represent it in detail   using $\zeta-$function
regularization. In the process, we found a uniform preservation of 
gauge invariance, under both small and large transformations, which in 
odd dimensions is linked with parity anomalies due to the possibility of 
spectral asymmetry. We thereby connected  the machinery of index  theorems
to the more prosaic question of how (``improved"!) Chern-Simons terms 
(that carry the large gauge information) could be present  in the 
finite temperature thermal field theory regime without violating 
the overall gauge invariance; this was closely related to the 
$\eta-$function. From our original representations, we were able to
give ``spectral'' representations  for the massive Dirac determinant
in terms  of the massless $\zeta$ and $\eta$ functions. In turn, this 
enabled us to provide explicit expansions for both parity even and odd 
parts of the effective action in the small as well as large mass limit.
A number of subtleties inherent in these expansions were discussed. One 
important aspect is that there is a finite regularization
ambiguity in the full non-perturbative action that 
parallels the well-known perturbative one where there is a residual 
Pauli-Villars regulator ambiguity:  results depend on the sign of its mass  
even as it  tends to infinity. For us, the ambiguity was in a twofold 
possibility of complex plane cut. A physically appealing choice was to insist 
on ``decoupling'' as the electron's mass 
becomes infinite. These ambiguities differ from the usual polynomial 
freedom associated with regularization,
simply because there are no gauge invariant polynomials available 
here. Instead,
they are reflected in the nearest possible way to that: through the
discrete value of the coefficient of $\eta(0)$, 
which contains the local, ``polynomial'' CS term.
In connection with the importance of the flat direction as representing the 
large gauge aspects, we  noted that  these aspects
would only  be found in perturbative  diagrams if one used fermion
propagators in the ``flat potential vacuum'' rather than simply the usual free 
ones. 
Finally,  we provided some explicit gauge field configuration examples
to show the emergence of our general results in  concrete cases involving  
external fields. 

This work is supported by NSF grant PHY-9315811, 
in part by funds provided by the U.S. D.O.E.
under cooperative agreement \#DE-FC02-94ER40818
and by INFN, Frascati, Italy. 

\appendix
\section*{A}

\renewcommand{\theequation}{A.\arabic{equation}}
\setcounter{equation}{0}

We start by considering $\Gamma_{PC}[A]$ defined in  (\ref{PCA}). Recalling
that
\begin{equation}
K_{\mp}(t,0)=\Phi\left(\frac{1}{2},\frac{1}{2},-m^2 t\right)=\exp(-m^2 t),
\end{equation}
the first term in  (\ref{PCA}) can be cast as
\begin{equation}
\frac{1}{2}\left.\frac{d}{ds}\zeta_{{D\!\!\!\!/}^{~2}+m^2}(s)\right|_{s=0},
\end{equation}
{\it i.e.} $1/2$ the effective action corresponding to the massive
Laplacian on the spinor. Because of the $1/\Gamma(s)$ factor in front 
of the integral,
the simple poles at $s=0$ of the second integral  in (\ref{PCA}) give rise to a
non-vanishing result. Since the singular behavior occurs when $t$ is near $0$
(the integral is regular near $t=\infty$), we can reduce the integration region
to the finite interval $(0,1)$ and use the asymptotic expansion for 
${\rm Tr}[\exp(-t{D\!\!\!\!/}^{~2})]$: 
\begin{equation}
{\rm Tr}[\exp(-t{D\!\!\!\!/}^{~2})]=\sum_{n=0}^{\infty} H_n
~t^{(n-d)/2},
\label{coeffo1}
\end{equation}
to evaluate the integral. Here $H_{n}$ are the Seeley-deWitt coefficients, 
local functionals of the gauge field and background geometry, invariant under 
small and large transformations. The integral turns out to be 
\begin{equation}
\label{ppp}
\mp \frac{\pi i}{2}\zeta_{{D\!\!\!\!/}^{~2}+m^2}(0)+\frac{1}{2}
\lim_{s\to 0}\frac{1}{\Gamma(s)}\int^1_0~dt~t^{s-1}\sum_{n=0}^{\infty} H_n
t^{{(n-d)}/{2}}\left[\frac{d K^{(1)}}{ds}\pm 2 m\sqrt{t}
\frac{d K^{(2)}}{ds}\right].
\end{equation}
Let us drop, for the moment, the contribution proportional to
$\zeta_{{D\!\!\!\!/}^{~2}+m^2}(0)$. Taylor-expanding $K^{(i)}$
in the second term of eq. (\ref{ppp}) and performing the integral
in $t$ we obtain
$$
\frac{1}{2}\lim_{s\to 0} \sum_{n=0}^{\infty}\sum_{k=0}^\infty\left [
\frac{H_n}{k!} \frac{d^k}{dt^k}\left(\frac{d}{ds}K^{(1)}(t,s)\right)_{t=0}
\frac{1}{\Gamma(s)\left(\displaystyle{s+\frac{n-d}{2}+k}\right)}
\right. +\nonumber\\
$$
\begin{equation}
\left. \pm 2 m\frac{H_n}{k!} \frac{d^k}{dt^k}
\left(\frac{d}{ds}K^{(2)}(t,s)\right)_{t=0}
\frac{1}{\Gamma(s)\left(\displaystyle{s+\frac{n-d+1}{2}+k}\right)}
\right ].
\end{equation}
Letting $s\to 0$ in the previous equation, because of the $1/\Gamma(s)$,
we will get a vanishing result unless $n=d-2k$ or $n=(d-1)-2k$. Thus we
can write
\begin{equation}
\frac{1}{2}\sum_{n=0}^{\infty}\sum_{k=0}^\infty\delta_{n,d-2k}
\frac{H_{n}}{k !} \frac{d^k}{dt^k}\left(\frac{d}{ds}K^{(1)}(t,s)
\right)_{s,t=0}\pm m\delta_{n,(d-1)-2 k}\frac{H_{n}}{k!} \frac{d^k}{dt^k}
\left(\frac{d}{ds}K^{(2)}(t,s)\right)_{s,t=0}.
\end{equation}
Taking into account of the fact that only the even coefficient $H_{2n}$ are 
different from zero in the heat-kernel expansion for the Laplacian, the first 
term contributes only  if $d$ is even while the second only if $d$ is odd.
Explicitly we have
\begin{eqnarray}
\label{pqpq}
\Gamma_{PC}[A]&=&\!\!\frac{1}{2}\left.\frac{d}{ds}
\zeta_{{D\!\!\!\!/}^{~2}+m^2} (s)
\right|_{s=0}\!\!\!\!\mp\frac{\pi i}{2}\zeta_{{D\!\!\!\!/}^{~2}+m^2}(0)+
\nonumber\\
&&+\ \ \left\{\begin{array}{l}
\displaystyle{\frac{1}{2}\sum_{k=0}^{d/2} \frac{H_{d-2 k}}{k!}
\frac{d^k}{dt^k}\left(\frac{d}{ds}
K^{(1)}(t,s)
\right)_{s,t=0}}\ \ \ \ \   d~~{\rm even}\\
\\
\displaystyle{\pm m\sum_{k=0}^{(d-1)/2} \frac{H_{d-1-2 k}}{k!}
\frac{d^k}{dt^k}\left(\frac{d}{ds}K^{(2)}(t,s)
\right)_{s,t=0}}\ \ \ \ \ \ \  d~~{\rm odd}\\
\end{array}\right.
\end{eqnarray}
where we have restored the contribution proportional to the
$\zeta(0)=\zeta_{{D\!\!\!\!/}^{~2}+m^2}(0)$. Let us notice
that the cut ambiguity affects  odd dimensions, while
the local terms in even dimension are insensitive to it. The
$\zeta(0)$ part can, as usual, be reabsorbed in a redefinition of
the scale. The local parts can be explicitly computed with the result 
\begin{equation}
\Gamma^{odd}_{Loc}[A]=\pm {\rm sign}(m)\sqrt{\pi}
\sum_{k=0}^{(d-1)/2} 
\frac{(-2)^{k} }{(2k+1)!!}(m^2)^{k+1/2} H_{d-1-2 k}
\label{coco1}
\end{equation}
in the odd dimensional case, while for $d$ even
\begin{equation}
\Gamma^{even}_{Loc}[A]=-\sum_{k=1}^{d/2} 
\left [\sum_{j=1}^{k} \frac{(-2)^{j-1}}{j}
\frac{1}{(2j-1)!!(k-j)!}\right ] (im)^{2k} H_{d-2 k}.
\label{coco2}
\end{equation}

\section*{B}

\renewcommand{\theequation}{B.\arabic{equation}}
\setcounter{equation}{0}

There is a very standard technique for  evaluating the derivative at  $s=0$ of 
a series such as
\begin{equation}
{\cal F}(s)=\sum_{\mu_k}\sum_{n\in Z\!\!\!Z}\left [\frac{\beta^2}{4\pi^2} 
(m^2+\mu^2_k)+\left (a+n+\frac{1}{2}\right)^2 \right]^{-s}.
\end{equation} 
One starts by writing a Mellin-representation of the complex power  and then
interchanges the sum  with the integral in $t$
\begin{eqnarray}
{\cal F}(s)&=&\sum_{\mu_k}\sum_{n\in Z\!\!\!Z}\frac{1}{\Gamma(s)}\int^\infty_0 dt ~t^{s-1}\exp
\left [ -\frac{\beta^2}{4\pi^2} (m^2+\mu^2_k) t-\left (a+n+\frac{1}{2}
\right)^2 t \right]=\nonumber\\
&=&\frac{1}{\Gamma(s)}\sum_{\mu_k}\int^\infty_0 dt ~t^{s-1} \sum_{n\in Z\!\!\!Z}
\exp \left [ -\frac{\beta^2}{4\pi^2} (m^2+\mu^2_k)t-\left (a+n+\frac{1}{2}
\right)^2 t\right].
\end{eqnarray}
The above integral exhibits a singularity at $s=0$ when $t$ approaches zero as well.
In order to remove this obstacle and thus compute ${\cal F}^\prime(0)$, we can use
Poisson-resummation, namely the identity
\begin{equation}
\sum_{n\in Z\!\!\!Z} f(n)=\sum_{n\in Z\!\!\!Z} \hat f(n),
\end{equation}
where $\hat f(n)=\displaystyle{\int_{-\infty}^{\infty} d x} \exp(2\pi i n x) f(x)$.  In our
case we have
\begin{equation}
{\cal F}(s)=\frac{\sqrt{\pi}}{\Gamma(s)}\sum_{\mu_k}
\int^\infty_0 dt ~t^{s-3/2}\sum_{n\in Z\!\!\!Z}
\exp \left [-\frac{\beta^2}{4\pi^2} (m^2+\mu^2_k) t -\frac{\pi^2 n^2}{t}-
2\pi i \left(a+\frac{1}{2}\right) n \right].
\end{equation}
Notice that the integral is now regular at $t=0$ for every $s$, when
$n$ is different from zero, so we can write
\begin{eqnarray}
&&{\cal F}(s)=\frac{\sqrt{\pi}}{\Gamma(s)}\sum_{\mu_k}\int^\infty_0 dt ~t^{s-3/2}
\sum_{n\in Z\!\!\!Z,~n\ne 0}
\exp \left [ -\frac{\beta^2}{4\pi^2} (m^2+\mu^2_k) t -\frac{\pi^2 n^2}{t} -
2\pi i\left(a+\frac{1}{2}\right) n \right]\nonumber\\
&&+\sqrt{\pi}\Gamma(s-1/2)\left [ \zeta_{\beta^2/4\pi^2({D\!\!\!\!/}^2+m^2)}(s-1/2)-
(\beta^2 m^2/ 4\pi^2)^{-s+1/2}(\nu_+ +\nu_-)\right ].
\end{eqnarray}
Performing the integral, we obtain
\begin{eqnarray}
&&{\cal F}(s)=\frac{\sqrt{\pi}}{\Gamma(s)}
 \sum_{\mu_k} \sum_{n\in Z\!\!\!Z,~n\ne 0}
\exp \left [ -2\pi i\left(a+\frac{1}{2}\right) n \right]
\left(\frac{2\pi^2 |n|}{\beta\sqrt{\mu^2_k+m^2}}\right)^{s-1/2}
\!\!\!\!\! K_{s-1/2}(\beta |n|\sqrt{\mu^2_k+m^2})\nonumber\\
&&+\Gamma(s-1/2)\sqrt{\pi}\left [
\zeta_{\beta^2/4\pi^2 ({D\!\!\!\!/}^2+m^2)}(s-1/2)-
(\beta^2 m^2/4\pi^2)^{-s+1/2} (\nu_+ +\nu_-)\right ],
\end{eqnarray}
where $K_\nu(x)$ is the  Bessel function.
We can now take the derivative and let $s\to 0$, because
the series is convergent and defines a holomorphic function at
$s=0$,
\begin{eqnarray}
{\cal F}^\prime(0)&=&
\sum_{n\in Z\!\!\!Z,~n\ne 0}\frac{1}{|n|}
\exp \left [ -2\pi i\left(a+\frac{1}{2}\right) n
-\beta |n|\sqrt{\mu^2_k+m^2}\right]\noindent\\
&&+2\pi\zeta_{({D\!\!\!\!/}^2+m^2)}(-1/2)-\beta m (\nu_+ +\nu_-).
\end{eqnarray}
Recalling that $\log(1-x)=\displaystyle{\sum_{k=1}^{\infty} 
\frac{x^k}{k}}$, we can compute the sum and finally find
\begin{equation}
{\cal F}^\prime(0)=
\log\prod_{\mu_k}\left |1+\exp\left[2\pi i a+\beta \sqrt{\mu_k^2+m^2}\right]\right |^2
+2\pi\zeta_{{D\!\!\!\!/}^2+m^2}(-1/2)-\beta m (\nu_+ +\nu_-),
\end{equation}
as reported in \cite{us}.
  
\end{document}